\input{aipcheck}

\documentclass[
    ,final            
  ]
  {aipproc}

\layoutstyle{6x9}

\begin{document}

\title{The role of environment in triggering starburst galaxies: A sample
of isolated galaxies}

\author{U. Lisenfeld}{
  address={Dept. de F\'\i sica Te\'orica  y del Cosmos, Universidad de Granada, Spain}
}

\author{L. Verdes-Montenegro}{
  address={Instituto de Astrof\'\i sica de Andaluc\'\i a, Camino Bajo de Hu\'etor 50, Granada, Spain}
}

\author{D. Espada}{
  address={Instituto de Astrof\'\i sica de Andaluc\'\i a, Camino Bajo de Hu\'etor 50, Granada, Spain}
}

\author{E. Garc\'\i a}{
  address={Instituto de Astrof\'\i sica de Andaluc\'\i a, Camino Bajo de Hu\'etor 50, Granada, Spain}
}

\author{S. Leon}{
  address={Instituto de Astrof\'\i sica de Andaluc\'\i a, Camino Bajo de Hu\'etor 50, Granada, Spain}
}
\author{J. Sabater}{
  address={Instituto de Astrof\'\i sica de Andaluc\'\i a, Camino Bajo de Hu\'etor 50, Granada, Spain}
}
\author{J. Sulentic}{
  address={Department of Physics \& Astronomy,
Tuscaloosa, AL 35487 USA}
}
\author{S. Verley}{
  address={Instituto de Astrof\'\i sica de Andaluc\'\i a, Camino Bajo de Hu\'etor 50, Granada, Spain}
,altaddress={LERMA, Observatoire de Paris, 61, Av. de l'Observatoire, 75014 Paris, France}}

\begin{abstract}
The project AMIGA (Analysis of the interstellar Medium of Isolated GAlaxies)
will provide a statistically significant sample of the 
most isolated galaxies in the northern sky.
Such a control sample is necessary to assess the role of the environment
in galaxy properties and evolution.
The sample is based on the Catalogue of Isolated Galaxies (CIG, Karachentseva, 1973)
and 
the database will include blue and near-infrared luminosities,
far-infrared (FIR) emission, atomic gas (HI) emission, radio continuum,
and, for a redshift limited subsample of about 200 galaxies, CO and H$\alpha$ emission.
The data will be released and periodically updated at
http://www.iaa.csic.es/AMIGA.html.
Here, we present the project and its status, as well as
a preliminary analysis of the relation between star formation
activity and the environment. 
We found a trend that the galaxies with asymmetric HI spectrum also  tend
to show an enhanced star formation, traced by a high value of
 $L_{FIR}/M(HI)$, indicating that a perturbed gas 
kinematics and enhanced SF are related.
\end{abstract}

\maketitle


\section{The link between environment and galaxy properties}

It is commonly believed that the environment has an influence on galaxy
properties and in particular on the star formation (SF) activity.
For example, all known ultraluminous infrared galaxies are mergers, which
shows that strong interactions can cause bursts of star formation. However,
such a relation is not obvious for weaker interactions. Whereas 
some authors find an increase of SF in interacting galaxies
(Larson \& Tinsley 1978, Bushouse 1987, Xu \& Sulentic 1991), 
other conclude that there is only a moderate increase
which is furthermore  restricted to the central regions of the
galaxies (Bergvall et al. 2003).
With respect to other galaxy properties the situation is similarly complex
and contradictory. For example, it is unclear whether the molecular gas content
is affected by gravitational interactions (see e.g. Braine \& Combes 1993,
Leon et al. 1998, Perea et al. 1997, Verdes-Montenegro et al. 1998,  for different results). 

This shows that although a relation between the environment and at least
some galaxy properties is present, we are far from understanding the link and
unable to quantify possible relations.
Many of the uncertainties are due to the lack of a suitable reference sample.
Such a sample should ideally include the most isolated galaxies and
consist of a multiwavelength database enabling us to characterize the properties
of star formation and the interstellar medium (ISM) in galaxies. 
In the project  AMIGA (Analysis of the interstellar Medium of Isolated GAlaxies)
our goal is to provide and analyse such a reference sample. In this contribution
we will give
a short description of the project and its status and present some preliminary results
with respect to the relation between star formation activity and environment.

\section{The AMIGA  project: A sample of the most isolated galaxies}

\subsection{The sample}

Our sample is based on the Catalogue of Isolated Galaxies (CIG, Karachentseva, 1973),
including 1050 galaxies. We chose this catalog because it presents several strengths:
(i) The large size of the catalog allows a statistically meaningful
analysis, even when refining the sample in isolation and morpholgy or restricting
the redshift range. (ii) The isolation criterium used by Karachentseva is clearly defined:
it contains galaxies for which
no similarly sized galaxies with diameter d, between
 1/4 and 4 times diameter D of the CIG galaxy, lie within 20d. Within this
definition, dwarf companions are not necessarily excluded.
(iii) All morphological types are found in the CIG, allowing an analysis as a function of
Hubble type. (iv) Finally, the CIG probes a large enough volume of space to allow us to sample
the majority of the optical luminosity function,
and it is 80-90\% complete to m$_{zw}= 15.0$ (Verdes-Montenegro et al. 2004).

\subsection{Refinement of the sample}

We are in the process of  
improving the CIG in several ways that take advantage of the
Digitized Sky Surveys (POSS1 and POSS2). 

\begin{itemize}

\item {\it Isolation revision:} We reevaluate the degree of isolation based
on red POSS1 images and a minimum physical diameter around each CIG galaxy of 0.5 Mpc
(Verley et al., in preparation).
The goals are to resolve close pairs, to identify missed and/or fainter
(candidate) companions, and to quantify the isolation degree.   
We used estimates of the tidal strengths and of the local density in order to
quantify the isolation degree. Both estimates give similar results. 
In the present paper, we will use the first estimator, 
defined as the ratio between the tidal force and binding force as an
estimate for the interaction strength:

\begin{equation}
Q=\frac{F_{tidal}}{F_{bind}} \propto \left(\frac{M_c}{M_p} \right) 
\left(\frac{d_p}{R}\right)^3 \propto
\frac{(d_c d_p)^{1.5}}{R^3}  \approx \frac{(D_c D_p)^{1.5}}{S^3} 
\end{equation}

where $M_c$/$M_p$ is the (unknown) mass of the companion galaxy/CIG galaxy 
which we assume to be related to the linear galaxy diameter  as
$M_{c/p} \propto d_{c/p}^{1.5}$ (Dahari 1984). 
$R$ is the distance between the galaxies.   Since 
 we do not have redshift information for most of
our candidate companions, we approximate $R$ by $S$, the projected distance, and
$d_{c/p}$ by the angular galaxy diameter $D_{c/p}$.  
In spite of the lack of the redshift information, $Q$
is expected to give a reasonable estimate of the tidal interaction strength in a 
statistical sense as can be seen from the following argument.
If the candidate companion galaxy is in reality a 
background object we have underestimated the true distance but also underestimated
the true size and mass. Both effects partly cancel out.
Only in the case of the candidate companion being a
foreground object $Q$ is overestimated.  Foreground objects are however rarer than
background objects because of the smaller volume sampled by them.
Nevertheless, it would be desirable to obtain the redshifts for the companion
galaxies and in the future we will try to achieve this at least for a subsample.

\item {\it Morphology revision:}
Galaxy classification data for CIG galaxies is nonuniform and often contradictory
when comparing e.g. NED vs. LEDA vs. CIG.
Based on POSS2 images it was possible to obtain reliable galaxy types for 
80\% of the sample (Sulentic et al., in preparation). 
The remaining 20\% of the sample are being supplemented with 
archival data or new CCD images on 1-2m class telescopes.

\item {\it Redshifts and distances:} 
We have searched the literature and archives for redshifts and distances of CIG galaxies,
revealing data for almost the entire sample (956/1050 galaxies).

\item {\it Positions:} 
The positions of the CIG galaxies were systematically revised
using SExtractor on images from the Digitized Sky Survey and visually checking
the results in cases of complex morphology. Differences between old and 
new positions of up to 38.9'' were found with a mean of 2.4'' (Leon \& Verdes-Montenegro 2003).

\end{itemize}

\subsection{The database}

The goal of this project is to build a multiwavelength database for the CIG that will allow us
to characterize the star formation activity and the properties of the different
components of the interstellar medium, study the interplay between ISM and SF
and the link between central activity 
(Seyfert, radio-galaxies) and  the environment.
In order to achieve this, we have 
obtained, either from archives, the literature or by own observations the
following data: 

\begin{itemize}

\item The optical luminosity as a tracer of the visible 
light and the stellar content. Zwicky magnitudes  are available for all CIG galaxies.
We have  corrected the Zwicky magnitudes for systematic errors reported by
Kron \& Shane (1976), Galactic dust extinction, internal dust extinction based on the
revised morphologies and applied a k-correction.

\item We plan to derive the near-infrared luminosity from 2MASS data, available for
90\% of the sample, as a measure of the 
old stellar population and a better tracer of the stellar mass.

\item H$\alpha$ as a good tracer
of recent star formation in places where extinction is not high.
H$\alpha$ images have  been taken for about 150 galaxies and the reduction is in progress.
These  images will allow us to quantify the total present SF
rate and to study its spatial distribution and morphology. 
 
\item  The far-infrared (FIR) emission, where most of the flux from 
newly formed stars is re-radiated. IRAS data is available for almost the 
entire sample. We have reprocessed the IRAS data using the tool
SCANPI which allowed us to increase the detection rate, improve the signal-to-noise
ratio for weak sources and derive more reliable fluxes for extended sources than the
Point Source Catalog. 

\item  Atomic gas, HI, as a fundamental ingredient of the ISM and
a very sensible tracer of  interaction. Data for about 800 galaxies
have been compiled partly from the literature and to a large extent 
from own observations with Effelsberg, Nancay and
Arecibo (Espada et al., in preparation).
The HI spectrum does not only provide the total mass of the atomic gas, but also allows us
to measure its asymmetry. We are currently testing  different parameters to 
quantify this asymmetry. In the present contribution,   
we use an asymmetry parameter
which is defined as the ratio of the velocity integrated 
intensities below and above the mean velocity.

\item CO emission as a tracer of the molecular gas (H$_2$) content, which 
represents the dense ISM and is the major building block for SF.
Very few CO observations of CIG galaxies could be found in the literature.
We have 
carried out observations for 183 galaxies with the IRAM 30m telescope, 
Five College Radio Telescope and the Nobeyama 45m telescope.

\item Radio-continuum emission as a useful tracer of
the current SF rate  and, through the deviation from the FIR-radio correlation,
for the nuclear activity. It has the advantage that it is not affected by extinction.
We reprocessed data from the VLA Sky Survey from which we obtained fluxes at 1.4 GHz
for 343 galaxies and from surveys of the Westerbrok
array (44 galaxies at 320 MHz) and from the Greenbank telescope (32 galaxies at 4.8 GHz) 
(Leon et al., in preparation).

\end{itemize}

It is impossible to carry out the  H$\alpha$ and CO observation for the whole
sample because it is too time-consuming. Therefore we have restricted these
observations to a redshift limited  sample (recession velocities between 1500 and
5000 km/s) of about 200 galaxies.
The data are being released and periodically updated at 
http://www.iaa.csic.es/AMIGA.html. 

\section{Star formation activity and the environment}

In the following, we use the FIR luminosity as a tracer of the SF rate.
Fig. 1 (left) shows the range of FIR luminosities of the sample spanning 
several orders of magnitude up to about $10^{11} L_\odot$.
Also the values of the FIR-to-blue ratio extend over 2 orders of magnitudes. 
We found a morphological segregation in this distribuion with irregular galaxies
being at the low luminosity end and 
early-type spirals at the high luminosity end.

We tested several ``starburst indicators'' commonly used in the literature
to trace galaxies with an  enhanced SF activiy.
Heckman et al. (1990) 
suggested the IRAS flux ratio  $S_{60}/S_{100} >  0.4$ as a tracer for
starburst galaxies. Dahlem et al. (2001) successfully used this tracer and
found it to correlate reasonably with the radio continuum luminosity per SF area. 
Rossa \& Dettmar (2003) found a similarly good correlation  for a sample
of edge-on galaxies between
 $S_{60}/S_{100}$ and  $L_{FIR}/D_{25}^2$.
We tested these parameters for our sample, as well as an additional one,
the FIR luminosity per atomic hydrogen mass, $L_{FIR}/M(HI)$.
In Fig. 1 (right) we show the comparision between two tracers,
$S_{60}/S_{100}$ and $L_{FIR}/M(HI)$ (with $L_{FIR}/D_{25}^2$ a similar relation
is found).
We only plot  data for galaxies with a signal-to-noise ratio $>$ 3 in
 both the 60 and 100 $\mu$m fluxes, in which case the maximum error
in  $S_{60}/S_{100}$ is a factor of 50\%.  
The relation between both quantities is slightly poorer than for the samples of
Dahlem et al. (2001) and Rossa \& Dettmar (2003), but still a trend is
visible. 
A possible reason for this poorer correlation  the relatively large measurement error in
$S_{60}/S_{100}$ in our sample with many weak IRAS sources.
Due to the small range spanned by $S_{60}/S_{100}$, this parameter 
 is  a useful starburst tracer
only for those galaxies with low  errors in the 60 and 100 $\mu$m data
so that even small differences in $S_{60}/S_{100}$ are significant.
In the following, we choose to use $L_{FIR}/M(HI)$ as an indicator for
enhanced SF. (Similiar results are  obtained with $L_{FIR}/D_{25}^2$ and,
restricting the sample to high signal-to-noise data, also with  $S_{60}/S_{100}$).

\begin{figure}
\centerline{\includegraphics[height=.3\textheight]{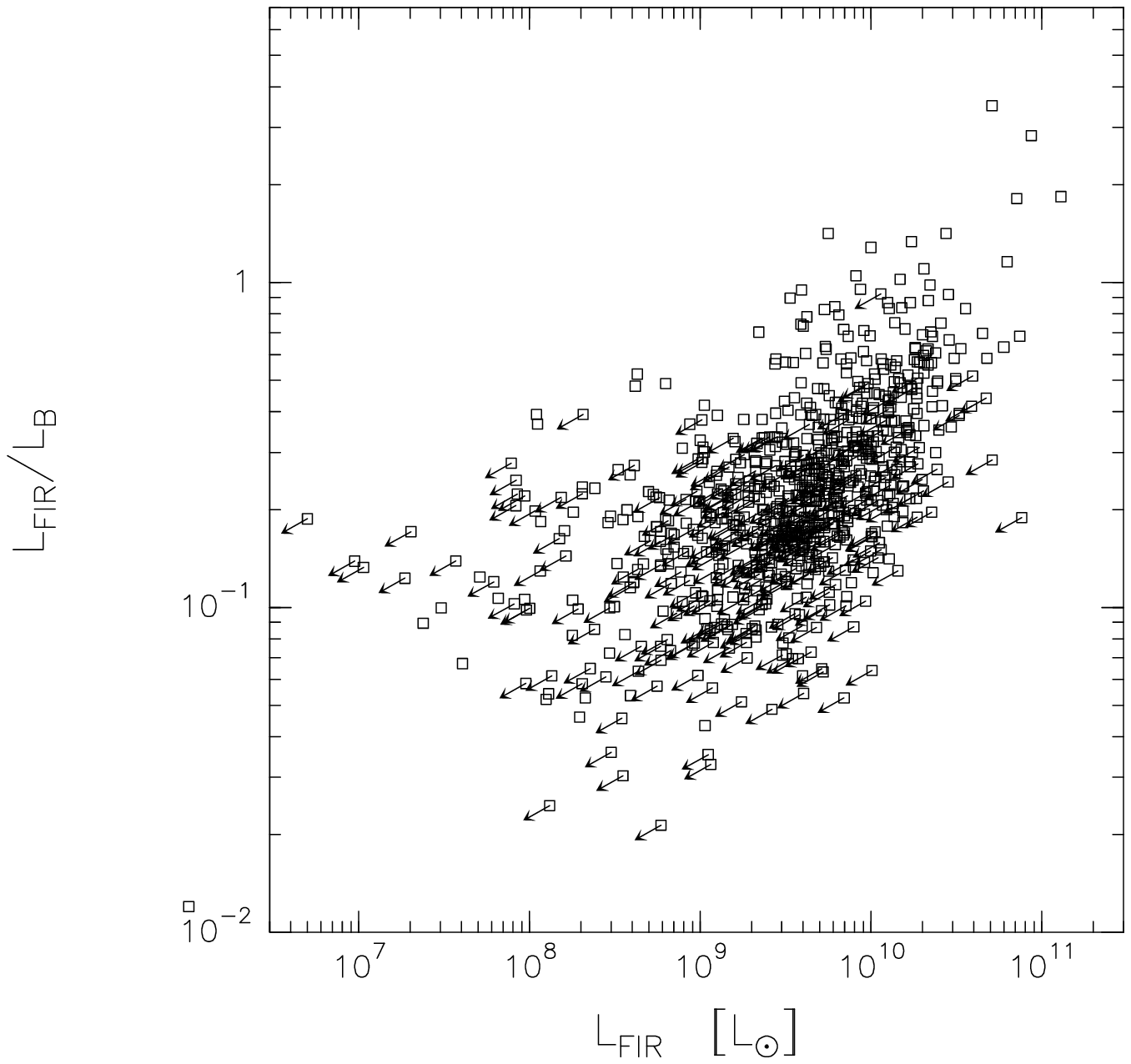}
\includegraphics[height=.3\textheight]{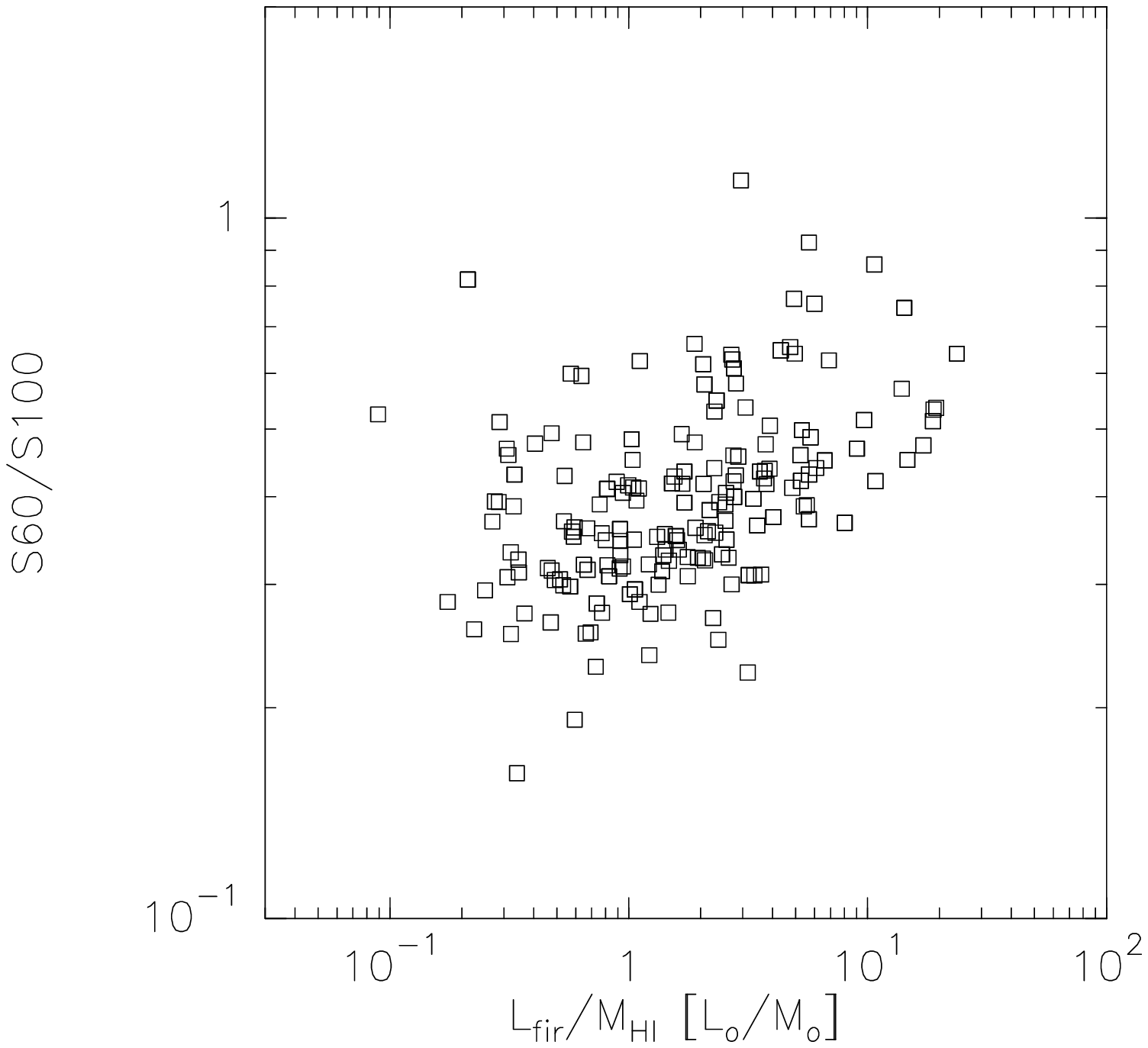}}
  \caption{{\bf Left:} FIR luminosity (calculated from the 60 and 100
$\mu$m fluxes) and FIR-to-blue luminosity ratio for the
of the sample with distance and IRAS data (n=293). {\bf Right:}
IRAS 60-to-100 flux ratio and FIR luminosity per HI mass for galaxies
with a signal-to-ratio $>$ 3 in both the 60 and 100 $\mu$m fluxes (n=175).
}
\end{figure}

We compared  $L_{FIR}/M(HI)$ to the tidal strength parameter $Q$, as defined in 
eq. (1). No correlation was found so far. However, this result might change in the 
future
with our refinement of this parameter, e.g. by taking into account the
redshift of candidate companions.

We do find, on the other hand, a trend when comparing $L_{FIR}/M(HI)$
to the asymmetry of the HI line.  Fig. 2 (left) shows 
that galaxies with the most asymmetric HI lines (i.e. those with values
strongly departing from 1 towards smaller and larger values) 
are those with high values of $L_{FIR}/M(HI)$. 
This is emphasized in Fig. 2 (right) where we plot the distribution of
the asymmetry parameter for galaxies with high and low  $L_{FIR}/M(HI)$.
The distribution is clearly much broader for galaxies with
$L_{FIR}/M(HI)\ge2$. Here, 43\% of the galaxies (34 out of 78) are
outside the central two bins, whereas for galaxies with $L_{FIR}/M(HI)<2$ 
only 13\%  (13 out of 97) lie outside.

\begin{figure}
\centerline{\includegraphics[height=.3\textheight]{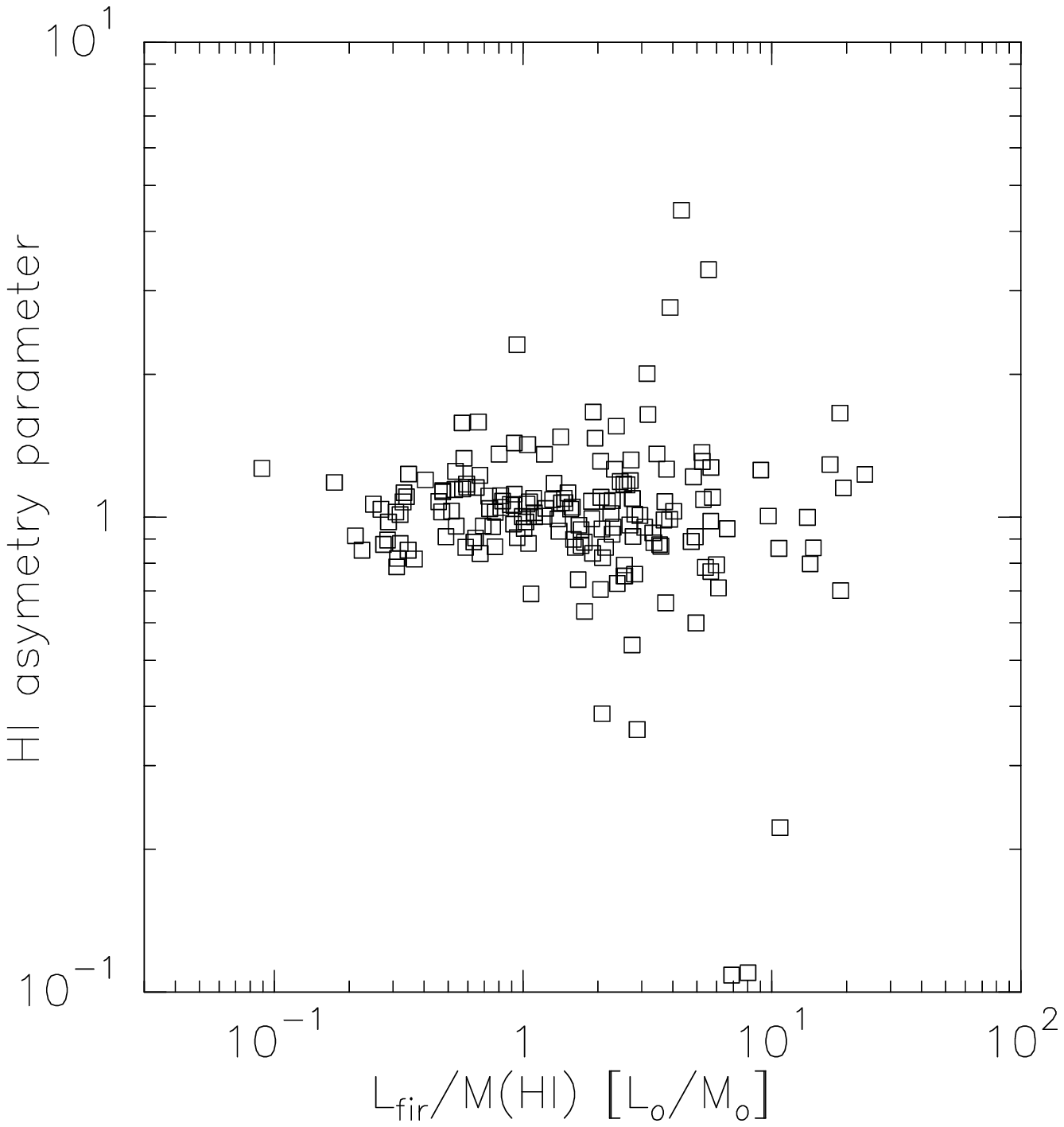}\quad
  \includegraphics[height=.3\textheight]{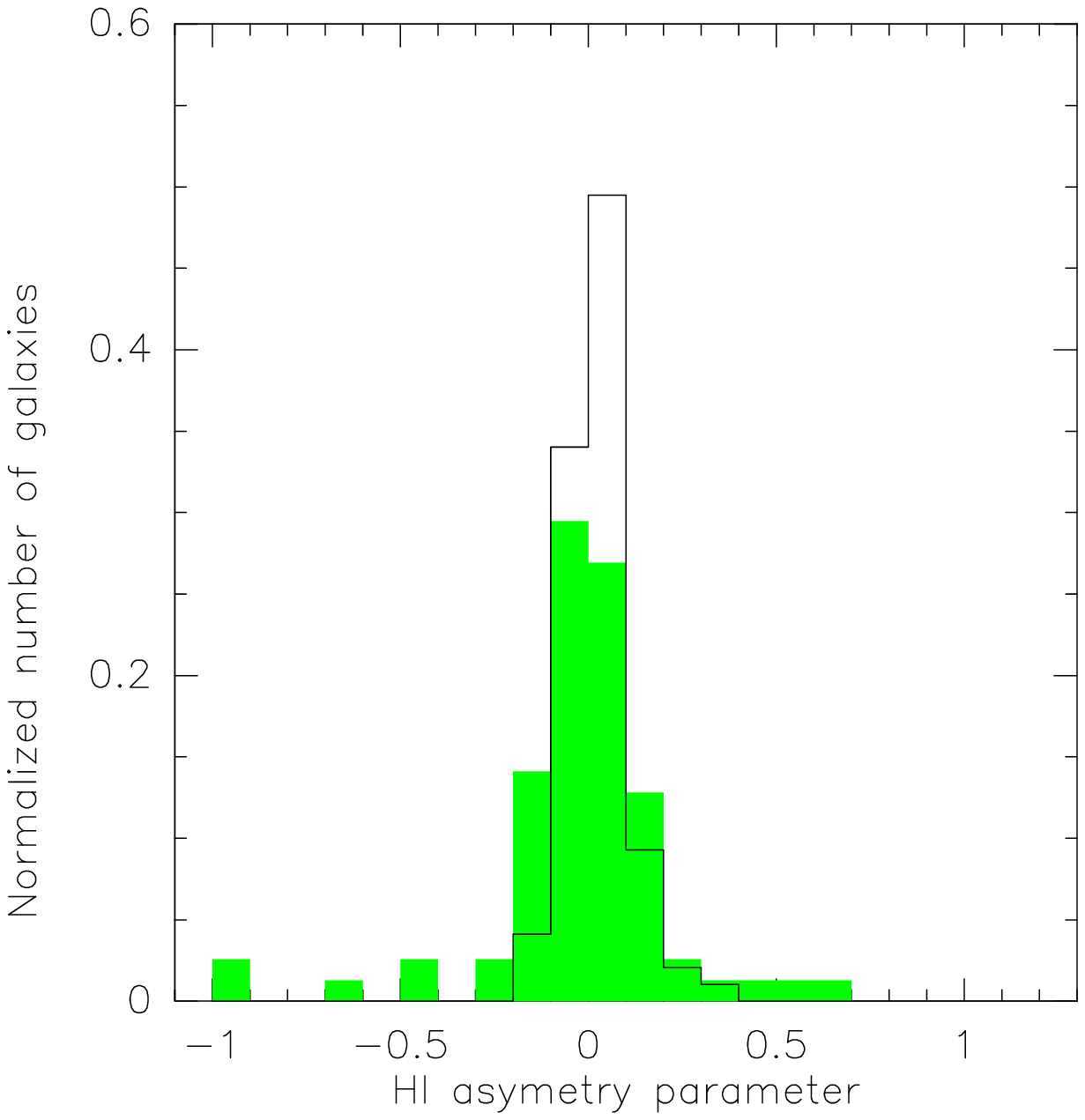}}
  \caption{{\bf Left:}  $L_{FIR}/M(HI)$ as a function of the
asymmetry parameter of the HI line, which is defined as the ratio of the velocity integrated intensities below and above the mean velocity giving values close to 1 for 
symmetric lines, and values  larger or smaller than 1 for asymmetric lines. 
{\bf Right:} Distribution of the
HI asymmetry parameter of the HI line for galaxies with
 $L_{FIR}/M(HI)\ge 2$ (shaded) and for $L_{FIR}/M(HI)< 2$ (black line). 
In both cases only galaxies with 
a signal-to-ratio $>$ 3 in both the 60 and 100 $\mu$m fluxes are included.
}
\end{figure}

The HI is a sensible  indicator for gravitational interaction.
The asymmetric HI spectrum could therefore hint at 
a tidal interaction with a weak companion not easily visible, or below the
sensitivity limit, in 
images of the Digitized Sky Survey.
However, the asymmetric HI spectrum does not prove that
an interaction is taking place. It could merely be an indicator for
instabilities in the disk -- whatever their causes are --
that produce an asymmetric gas distributions and kinematics
and might at the same time drive gas 
to the center of galaxies and enable an increase in the SF rate (see Combes, this volume).





\bibliographystyle{aipproc}   


\IfFileExists{\jobname.bbl}{}
 {\typeout{}
  \typeout{******************************************}
  \typeout{** Please run "bibtex \jobname" to optain}
  \typeout{** the bibliography and then re-run LaTeX}
  \typeout{** twice to fix the references!}
  \typeout{******************************************}
  \typeout{}
 }

\end{document}